\documentclass[%
 reprint,
 amsmath,amssymb,
 aps, pra
]{revtex4-2}

\usepackage{graphicx}% Include figure files
\usepackage{dcolumn}% Align table columns on decimal point
\usepackage{bm}% bold math
\usepackage{pbox}
\usepackage{hyperref}% add hypertext capabilities
\begin{document}

\preprint{APS/123-QED}

\title{Identifying influential node groups in networks with core-periphery structure}% Force line breaks with \\

\author{Gyuho Bae}
\email{gyu.bae@strath.ac.uk}
\affiliation{Department of Mathematics and Statistics, University of Strathclyde, Glasgow G1 1XQ, United Kingdom}
\author{Philip A. Knight}%
\affiliation{Department of Mathematics and Statistics, University of Strathclyde, Glasgow G1 1XQ, United Kingdom}
%\affiliation{Department of Mathematics and Statistics, University of Strathclyde\\ Glasgow, G1 1XQ, UK
%}%

% \collaboration{MUSO Collaboration}%\noaffiliation

\author{Young-Ho Eom}%
 %\homepage{https://sites.google.com/view/csns-lab/home}
 \email{yheom@uos.ac.kr}
 \affiliation{Department of Mathematics and Statistics, University of Strathclyde, Glasgow G1 1XQ, United Kingdom,}
 \affiliation{Department of Physics, University of Seoul, Seoul 02504, Republic of Korea,}
 \affiliation{Natural Science Research Institute, University of Seoul, Seoul 02504, Republic of Korea,}
 \affiliation{Urban Big data and AI Institute, University of Seoul, Seoul 02504, Republic of Korea}

% \affiliation{
%  Third institution, the second for Charlie Author
% \affiliation{%
%  Authors' institution and/or address\\
%  This line break forced with \textbackslash\textbackslash
% }%

% \collaboration{CLEO Collaboration}%\noaffiliation

\date{\today}% It is always \today, today,

%This contradiction implies that an expectation might be dependent on certain conditions.

\begin{abstract} 
Identifying influential spreaders is a crucial problem for practical applications in network science. The core-periphery (C-P) structure, common in many real-world networks, comprises a densely interconnected group of nodes (core) and the rest of the sparsely connected nodes subordinated to the core (periphery). Core nodes are expected to be more influential than periphery nodes generally, but recent studies suggest that this is not the case in some networks. In this work, we look for mesostructural conditions that arise when core nodes are significantly more influential than periphery nodes. In particular, we investigate the roles of the internal and external connectivity of cores in their relative influence. We observe that the internal and external connectivity of cores are broadly distributed, and the relative influence of the cores is also broadly distributed in real-world networks. Our key finding is that the internal connectivity of cores is positively correlated with their relative influence, whereas the relative influence increases up to a certain value of the external connectivity and decreases thereafter. Finally, results from the model-generated networks clarify the observations from the real-world networks. Our findings provide a structural condition for influential cores in networks and shed light on why some cores are influential and others are not.

% \begin{description}
% Secondary publications and information retrieval purposes.
% You may use the \texttt{description} environment to structure your abstract;
% use the optional argument of the \verb+\item+ command to give the category of each item. 
% \end{description}
\end{abstract}

\maketitle

%\tableofcontents

\section{\label{sec:Intro} Introduction}

Identifying influential spreaders in networks is relevant to crucial real-world problems such as capturing effective targets for vaccination or advertisement\cite{montes2022analysis, meng2022analysis, zhang2022sei3r, zareie2019influential}. An important problem is locating influential nodes in a network by leveraging its structure. The idea underlying structure-based identification of influential nodes is that a node's connectivity to the rest of the network determines its ability to spread (i.e., influence). Many methods have been proposed to identify influential nodes based on structural features such as degree, betweenness, and h-index, to name a few\cite{ma2016identifying, liu2016identify, lu2016vital, lu2016h, hu2013new, namtirtha2021best}. 

Core-periphery (C-P) structure\cite{borgatti2000models, holme2005core, csermely2013structure, kojaku2018core} is common in many real-world networks from the worldwide air-transportation network (WAN)\cite{verma2014revealing} to the Internet\cite{carmi2007model} and underground transportation networks\cite{lee2014density}. The core of a network is usually defined as a group of strongly connected nodes that are relatively well connected to the other part of the network. The other part, called the periphery, is a group of weakly connected nodes\cite{borgatti2000models, holme2005core, rombach2014core, csermely2013structure}. For instance, the WAN comprises well-connected hub airports (core) and less connected local airports (periphery) whose destinations are normally the hub airports. By the definition of the C-P structure, core nodes are expected to have more influence than periphery nodes. A seminal study\cite{kitsak2010identification} suggested that where a given node is located in the spectrum between the periphery and core is more crucial than its local connectivity (i.e., the number of its nearest neighbours in the network) in terms of node influence. Other work\cite{bae2014identifying, liu2015improving, zhang2016identifying, koujaku2016dense, malliaros2016locating, wang2017ranking} has highlighted the importance of the node location in terms of its influence as well.

However, a recent study\cite{liu2015core} has shown that core nodes are not always more influential than periphery nodes, in particular, when networks have community-like, isolated cores. Even if core nodes can have high local influence on nodes inside of the core, they can have low global influence outside of the core if the core has weak connectivity to the rest of the network. This result suggests that the influence of core nodes may depend on the internal and external connectivity of the core, and being in the core is not a sufficient condition for being an influential node. Although real-world networks exhibit a variety of the C-P structure\cite{zhang2015identification, kojaku2018core, gallagher2021clarified}, it is not well understood when cores are significantly more influential than their periphery.

In this study, we investigate how the influence of the core varies depending on its internal and external connectivity in real-world and model-generated networks. To do this, we consider two quantities: the internal connectivity of the core and the external connectivity of the core to the periphery. The internal connectivity quantifies how well the core nodes are connected to each other. The external connectivity quantifies how well the core nodes are connected to the periphery nodes. Our investigation of the C-P structure among 397 real-world networks shows that both internal and external connectivity are broadly distributed, indicating that the core's relative influence in these networks may also be broadly distributed. We estimated node influence with an extensive simulation of an epidemic model on 132 real-world networks and observed that the relative influence of the cores is broadly distributed in these networks. More specifically, we observed that the cores are significantly more influential than their periphery in only 78 real-world networks. The cores of the remaining 54 networks had comparable influence to their periphery, albeit slightly higher. We observed a strong positive correlation between the internal connectivity of a core and its relative influence but an optimal behaviour between the external connectivity and the relative influence of the core, where the relative influence increases up to a certain value but decreases thereafter. Tests on the 2-block-based model network clarify the results we observed in the real-world networks. Our findings indicate when the core of a network can be significantly more influential than the periphery and suggest that, if leveraging core-periphery structure to identify influential groups of nodes, it would be helpful to first look at the internal and external connectivity of the core.

\section{\label{sec:Method I} Evaluation of C-P structure and measures of structural features}

\subsection{Lip algorithm}

We use the Lip algorithm\cite{lip2011fast} to identify the core and periphery nodes of a given network. The algorithm is widely used\cite{ bargigli2015multiplex, lux2015emergence, barucca2016centrality, karlovvcec2016core} for detecting the network's C-P structure and is fast enough to identify the structure in a reasonable time. 

The $k$-shell decomposition algorithm\cite{alvarez2005k}, used in previous studies\cite{kitsak2010identification, liu2015core, malliaros2016locating}, assigns each node a $k$-shell number that denotes the location of the node from the outermost periphery to the innermost core. Since the maximum shell number can differ significantly between networks, a comparison between networks is difficult. A cut-off for the shell number is also necessary to decide a partition into a core and periphery, which can cause difficulty in identifying a distinct core. To avoid the above difficulties, we use the Lip algorithm for the identification of the core and periphery.

The Lip algorithm defines a set of core nodes and a complementary set of periphery nodes. The algorithm minimises an objective function $Z$ to find a core set. The objective function penalises links between two periphery nodes, while rewarding connections between two core nodes or one core node and one periphery node. The identification result from the Lip algorithm is different from other partitions based on centrality measures or other criteria\cite{alvarez2005k, lu2016h}. In the Lip algorithm, the objective function is 

\begin{align}
    Z(S_{1}) = \sum_{(i<j) \in S_{1}} \Pi_{(A_{ij} = 0)} + \sum_{(i<j) \notin S_{1}} \Pi _{(A_{ij} = 1)}
    \label{eqn: Lip method}
\end{align}

\noindent where $A_{ij}$ is $1$ if there is a link between node $i$ and $j$ or $0$ otherwise. The predication function $\Pi_{A_{ij} = 0\: \text{or}\: 1}$ returns $1$ if the condition $A_{ij} = 0\: \text{or}\: 1$ is true and $0$ if it is false. Thus, if a test set $S_{1}$ comprises well-connected members and the complementary set has sparsely connected members, both terms of the right-hand side of Eq.~\ref{eqn: Lip method} are minimised. This is the key feature of the algorithm to prevent falsely investigating nodes with large degrees that have sparser interconnections between them. For instance, ten interconnected nodes whose degree is $20$ can be identified as the core, whereas a node whose degree is $30$ can not be identified as the core when its neighbours have low degree.

\subsection{Structural features: link density}

 The characteristics of the C-P structure of a given network can be quantified by the link density in the core, in the periphery, and between the core and periphery because the main features of the C-P structure are a densely connected core, a sparsely connected periphery, and intermediate connectivity between the core and periphery. We can express the density of the core by the total number of links between core nodes($L_{cc}$) divided by the maximum capacity of the core($N_{c}^2 -N_{c}$) where $N_{c}$ denotes the number of nodes in the core. The same idea can be used for the periphery block and the block representing the external connectivity between the two partitions. Thus, we can define three link densities

\begin{align}
    \rho_{cc} &= 2L_{cc}/(N_{c}^2 -N_{c}),  \nonumber \\
    \rho_{cp} &= L_{cp}/(N_{c}N_{p}),  \nonumber \\ 
    \rho_{cp} &= 2L_{pp}/(N_{p}^2 - N_{p}),
    \label{eqn: link-densities-basic}
\end{align}

\noindent where $L_{ij}$ is the number of links in the corresponding block and $N_{i}$ is the number of nodes in the core or periphery. Since the Lip algorithm fulfils the C-P properties by identifying two groups of the dense and the sparse block, these link densities can characterize the C-P structure of networks.

However, these densities are not suitable for comparing different networks of different sizes. In this work, we define the internal connectivity as $\tilde{\rho}_{cc} = \rho_{cc}/\rho_{pp}$ and the external connectivity as $\tilde{\rho}_{cp} =\rho_{cp}/\rho_{pp}$, respectively. These ratios are helpful in seeing the details of the identified C-P structure\cite{urena2023assortative}. We note that the C-P structure in the ideal sense meets a criterion $\tilde{\rho}_{cc} > \tilde{\rho}_{cp} \gg 1$, while the isolated core reported in \cite{liu2015core} can be represented when $\tilde{\rho}_{cp} < 1$. $\tilde{\rho}_{cc} \approx 1$ and $\tilde{\rho}_{cp} \approx 1$ means the C-P structure's clarity is not higher than a random structure such as the Erdős–Rényi random network.

\section{\label{sec:Method II} node influence and core(periphery) influence}

\subsection{Evaluation of node influence}
  
We use Susceptible-Infected-Recovered(SIR) epidemic dynamics\cite{pastor2015epidemic}, a basic model for epidemic spreading, to estimate the influence of nodes in both model and real-world networks. In the SIR model, a node in a given network can be in one of three states: \textbf{S}usceptible, \textbf{I}nfected, and \textbf{R}ecovered. An infected node transmits the disease to one of its susceptible neighbours with probability $\beta$ and recovers spontaneously with probability $\gamma$. The recovered node can not be reinfected and remains recovered. The process ends when there are no infected nodes in the network. We define $R_{i}$ as the number of recovered nodes at the end of a spreading process started from node $i$. The influence of the node $i$ is $r_{i} = R_{i}/N$, where $N$ is the total number of nodes in the network\cite{kitsak2010identification, liu2015core, lu2016h}. 

Finding an influential node means that there must be differences in the influence of nodes. Therefore, finding a suitable $\beta$ for each network is crucial in this study. If $\beta$ is too large, the epidemics spread throughout the network regardless of the structure\cite{min2018identifying}. On the other hand, if $\beta$ is too small, only local spreading is possible, which is unsuitable for investigating a node's ability to act as a global spreader. Accordingly, we take an infection rate that means the most influential node is expected to infect at least $10 \%$(denoted as $\beta_{>10\%}$) of the network, according to previous studies\cite{kitsak2010identification, liu2015core, lu2016h}.

Since there is no prior information on $\beta_{>10\%}$ in general, we work with $\beta_{c} = \langle k \rangle /\left( \langle k^2 \rangle - \langle k \rangle \right)$, which is the epidemic threshold from the degree-based mean-field approach\cite{pastor2015epidemic}. It is known that there is a global epidemic if $\beta > \beta_{c}$ where a network's degree-degree correlation is $0$. Yet, the actual quantity of $\beta_{c}$ in real-world networks can be different from the $\beta_{c}$ in the networks with no correlation, and it does not necessarily coincide with $\beta_{>10\%}$. Thus, we pick $\beta$ from the set $\{ \beta_{c} /16, \beta_{c} /8, \beta_{c} /4, \beta_{c}, 4 \beta_{c}, 8 \beta_{c}, 16\beta_{c} \}$. The recovery rate $\gamma$ is set to $1$. If the initially infected node fails to infect any other nodes, then the spreading process will end immediately.

%If the first infected node $i$ is not in the core, the spreading process will end quickly since the recovery rate ensures immediate removal of the infected node.

%This condition also relates to the core's size for our model network, which allows a node in the isolated core block to have enough influence for the infection rate criterion. 

\subsection{The average influence of the core and periphery}

We can define the average value of influence for core and periphery(written $r_{c}$ or $r_{p}$) where

\begin{equation}
    r_{S_{c(p)}} = \frac{\sum_{i \in S_{c(p)}}r_{i}}{{N_{S_{c(p)}}}}, \quad S_{c(p)} = \text{core(or periphery)}.
    \label{eqn: ratio_influence}
\end{equation}

We can use $r_{c}/r_{p}$ to measure the relative difference in influence. We consider that if $r_{c}/r_{p} > 2$, then the core is significantly more influential than the periphery.

\begin{table}
\caption{\label{tab: terms}%
Descriptions of measures in this paper
}
\begin{ruledtabular}
\begin{tabular}{c|c} 
\colrule
 Quantity & Notation  \\ 
 \hline
 \# of total nodes & $N$  \\ 
 \hline
 \# of total links & $L$ \\ 
 \hline
 \# of nodes in core or periphery & $N_{c}, \: N_{p}$ \\ 
 \hline
 \# of links in blocks & $L_{cc},\: L_{cp}, \:L_{pp}$\\ 
 \hline
 Link density of blocks & $\rho_{cc},\: \rho_{cp},\: \rho_{pp}$ \\
 \hline
 Influence of node $i$ & $R_{i}$ \\
 \hline
 Normalised influence of node $i$ & $r_{i} = R_{i}/N$ \\
 \hline
 \pbox{20cm}{Average influence of \\ the core or periphery} & $r_{S} = \frac{\sum_{i \in S}r_{i}}{N_{S}},\: S = \text{C, P}$ \\
 \hline
 internal connectivity & $\tilde{\rho}_{cc} = \rho_{cc}/\rho_{pp}$ \\
 \hline
 external connectivity & $\tilde{\rho}_{cp} = \rho_{cp}/\rho_{pp}$ \\
\end{tabular}
\end{ruledtabular}
\end{table}

%For instance, using E-R model with three different parameters for $\rho_{ij}$      

\section{\label{sec: IV}Generating 2-block random network model}

%모델을 쓰는 이유를 나타내는 문장일텐데 모호합니다. "실제 네트워크는 C-P외에도 다양한 구조적 특징이 있으니 우리는 C-P외에는 사실상 무작위인 모델을 고려해서 C-P 구조가 노드 영향력을 어떻게 결정하는지 알아보고자 한다."라고 하는 게 어떨까요?

%Since real-world networks essentially hold complex characteristics more than C-P structure, using a simple enough model with a pre-defined C-P structure to understand how the detail of the C-P structure affects the node influence. 

Since real-world networks may show other structural features besides the C-P structure, we can clarify the role of the C-P structure in node influence if we consider model networks in which all other structures are random except for the controllable C-P structure. In this study, we apply a random network model with two stochastic random blocks of different link densities. This model has five parameters $(N,\alpha, \rho_{cc},\rho_{cp},\rho_{pp}$) for the generated networks. We adopt the Erdős–Rényi random network\cite{erdos1960evolution} as the null model. We set the network size as $N$ and the core size as $\alpha N$, where $ 0 < \alpha < 1$ denotes the proportion of the core. After setting the sizes of blocks and the average degree of the null model, the parameters $(\rho_{cc}, \rho_{cp}, \rho_{pp})$ determine each block's link density. The linking probability of two arbitrary nodes depends on those link densities. For instance, if two nodes are chosen from the core, the linking probability is equivalent to $\rho_{cc}$. Other cases for two chosen nodes from the periphery or one from the core and one from the periphery follow $\rho_{pp}$ and $\rho_{cp}$ as the linking probability, respectively. Note that the degree distributions of generated networks based on a combination of Poisson distributions differ greatly from those of real-world networks. 

We investigate the relations between $r_{c}/r_{p}$ and $(\tilde{\rho}_{cc}, \: \tilde{\rho}_{cp})$ in the model-generated and real-world networks to reveal the difference in influence between core and periphery nodes in networks with C-P structure.

\section{\label{sec:Method V}Data and algorithm}
We picked 397 real-world networks from network collections\cite{Netzsch, konect} that are categorised in many criteria, such as affiliation, animal interaction, and human contact, to name a few. The size of sampled networks is distributed from $\sim 30$ to $\sim 10^{5}$, and the maximum size of the simulated real-world networks is $\sim 8.5\times 10^{5}$. The average degree of networks broadly distributed from $\sim 1$ to $\sim 80$. We remove the directions and weight of links in the considered networks to avoid too much complexity.

\section{\label{sec:Result I}Results}

\subsection{Relationship between the core's connectivity and core influence in real-world networks}

\begin{figure}[ht]  

  \centering
      \includegraphics[width=\linewidth]{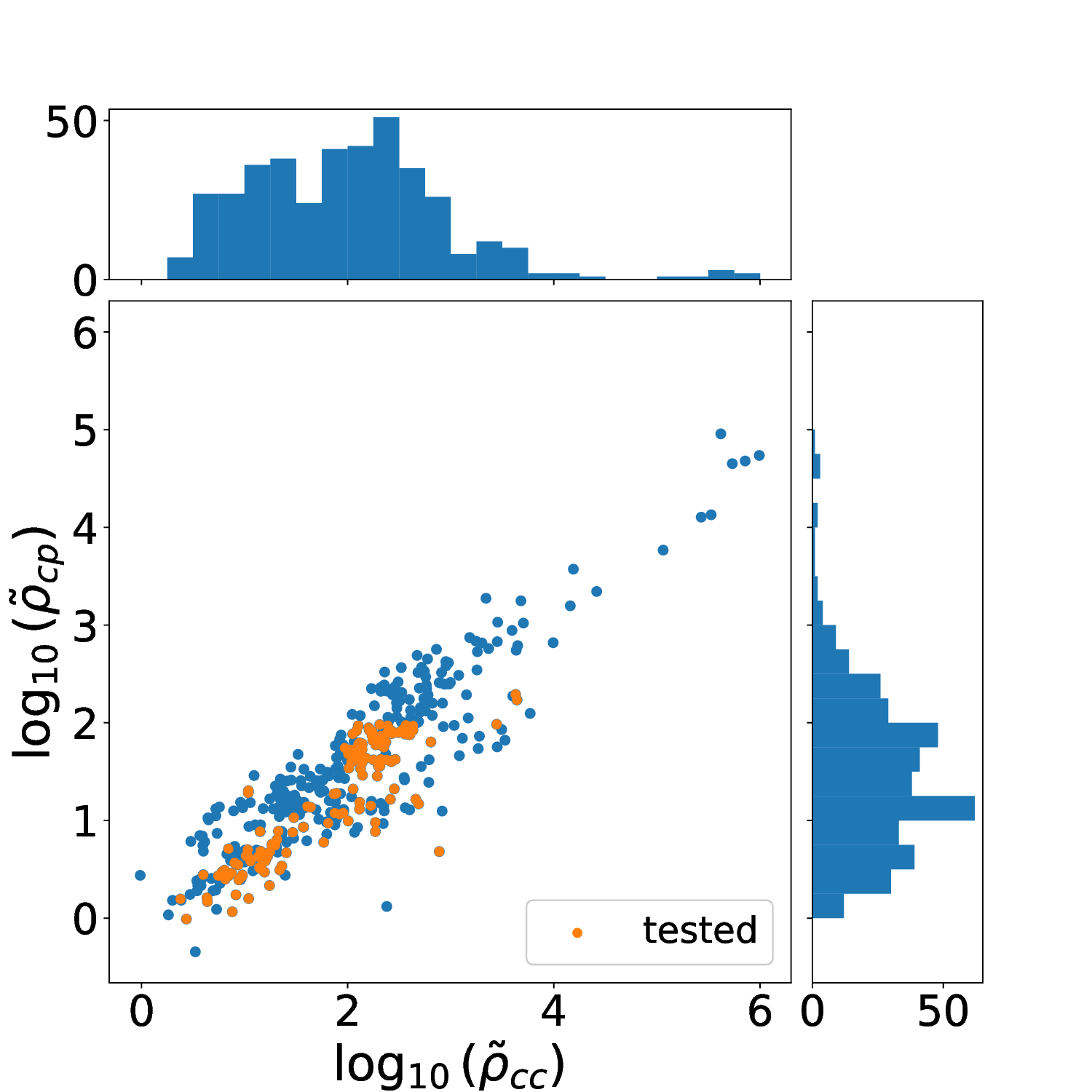}
  \caption{Observed internal and external connectivity$(\tilde{\rho}_{cc}, \:\tilde{\rho}_{cp})$ of the C-P structure of 397 real-world networks. Each histogram represents the frequency of networks in the range of values. Orange dots represent 132 networks that have been selected for SIR simulations.}
  \label{fig:corr_cla_intr}

\end{figure}

\begin{figure}[ht]  

  \centering
      \includegraphics[width=\linewidth]{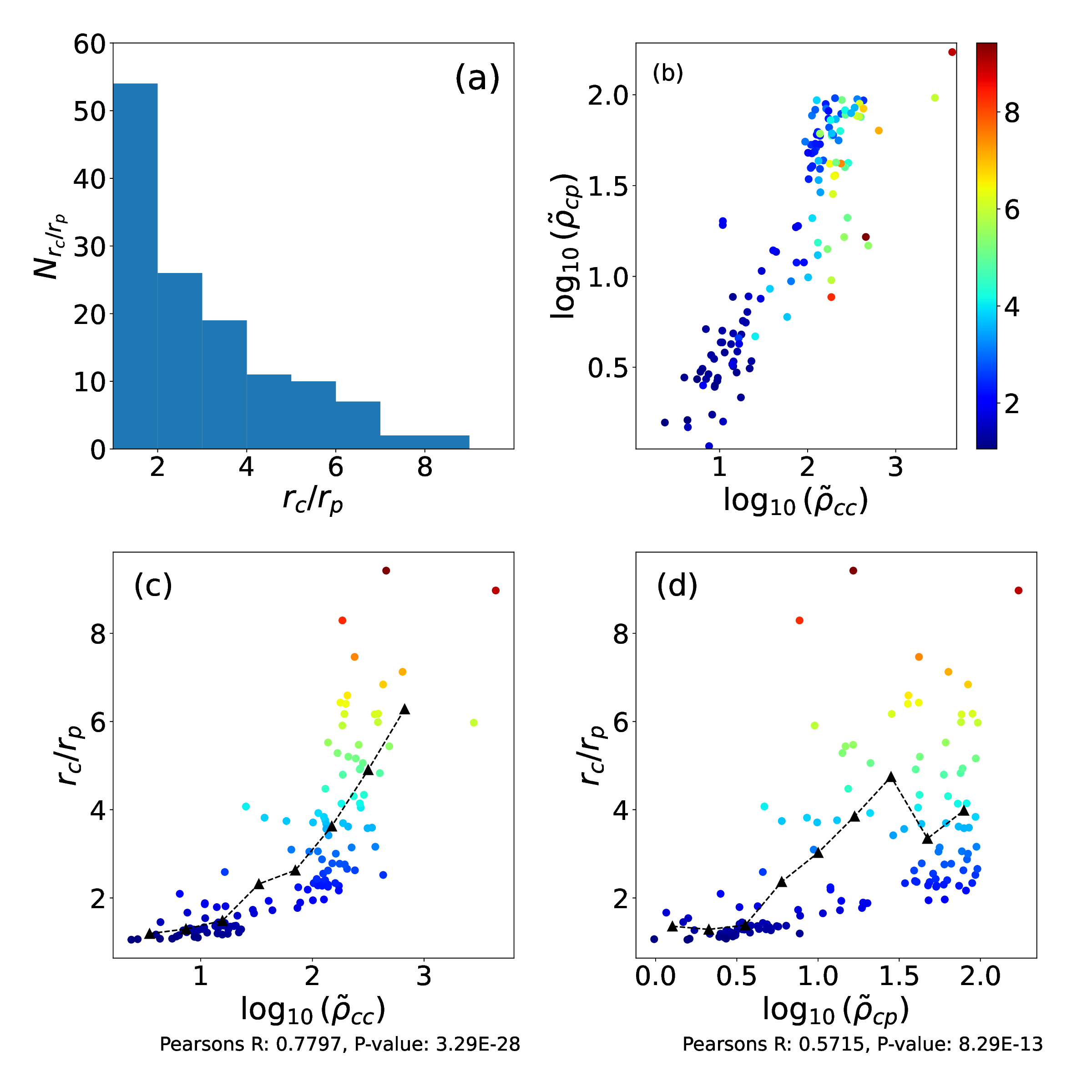}
  \caption{The core's relative influence $r_{c}/r_{p}$ and its relationship with $(\tilde{\rho}_{cc}, \:\tilde{\rho}_{cp})$ of 132 tested real-world networks. FIG~\ref{fig:influence_pattern_real-world}. (a) demonstrates the histogram of $r_{c}/r_{p}$ for 132 tested networks. FIG.~\ref{fig:influence_pattern_real-world}. (b) represents the relationship between $r_{c}/r_{p}$ and the ratio of two structural features. FIG.~\ref{fig:influence_pattern_real-world}. (c) and (d) represents the relationship between $r_{c}/r_{p}$ and the connectivity of the cores to their network($\tilde{\rho}_{cc}$ and $\tilde{\rho}_{cp}$), respectively. The dashed line with triangle markers in FIG.~\ref{fig:influence_pattern_real-world}. (c) shows averages of $r_{c}/r_{p}$ within $\log_{10}(\tilde{\rho}_{cc}) \pm 0.5B$, where $B = \frac{max(\log_{10}(\tilde{\rho}_{cc})) - min(\log_{10}(\tilde{\rho}_{cc}))}{10}$ is the width of sampling column. There are only two data points with $log_{10}(\tilde{\rho}_{cc}) > 10^3$, and we have excluded them as outliers. This line is applied in the same manner for (d). Each legend shows the Pearson correlation coefficient between  $r_{c}/r_{p}$ and corresponding connectivity. In FIG.~\ref{fig:influence_pattern_real-world}. (b) $\sim$ (d), axes are denoted as decimals, but the quantity is logarithm base 10. The colour of the dots in those plots indicates the values of $r_{c}/r_{p}$.}
  \label{fig:influence_pattern_real-world}

\end{figure}

%[1] intraconnectivity(internal connectivity), external connectivity 가 core influence 에 중요한 역할을 할 것으로 보이니 약 400 개의 네트워크에 대해 두 값들을 조사, 분포가 다양함을 확인(다양함이란 broadly distributed) 그 중에서도 어느 영역이 많다 라는 서술을 해야 함. correlation 에 대한 설명은 어떤 정보를 가지지는 않음. 

Identifying cores in networks can be helpful for finding influential nodes. However, the level of the relative influence of the identified cores may depend on their internal and external connectivity. 

First, we investigated the internal and external connectivity of cores in real-world networks. To do this, using the Lip algorithm, we identified a core and a periphery in each of 397 real-world networks. Then, we calculated the internal and external connectivity in each of the real-world networks. Both distributions of the internal and external connectivity are broad (across several orders of magnitude), indicating a rich diversity in the C-P structure of the real-world networks.

The histograms are at the top and at the right in FIG.~\ref{fig:corr_cla_intr} show the distributions of the internal connectivity($\tilde{\rho}_{cc}$) and external connectivity($\tilde{\rho}_{cp}$), respectively. They are distributed across order of magnitude from $10^{0}$ to over $10^{6}$, and mainly placed between $10^{0}$ and $10^{4}$.

FIG.~\ref{fig:corr_cla_intr} shows that $\tilde{\rho}_{cc}$ is generally greater than $\tilde{\rho}_{cp}$, which means network cores are generally denser than the interconnections between partitions. However, we can see that there are some networks with $\tilde{\rho}_{cc} < \tilde{\rho}_{cp}$. These do not have a pronounced C-P structure as the ideal definition requires $\tilde{\rho}_{cc} \geq \tilde{\rho}_{cp} \gg 1$. The isolated core($\tilde{\rho}_{cp} < 1$), observed in \cite{liu2015core}, is rarely observed in identified C-P structures with the Lip algorithm, but two networks, one of social interactions between zebras and another concerned with human social interactions, can be fitted in this category. 

Second, we investigated the relationship between the two connectivities($\tilde{{\rho}}_{cc}, \:\tilde{{\rho}}_{cp}$) and the relative influence of the core $r_{c}/r_{p}$ in real-world networks. To do this, we randomly selected 132 networks since the high computational cost of the SIR model simulations make a comprehensive test impractical.

The result in FIG.~\ref{fig:influence_pattern_real-world}. (a) shows a substantial number(54) of network cores are not significantly influential ($r_{c}/r_{p} \leq 2$) than their periphery. This means that there are relatively less influential cores, which makes finding influential nodes inside the cores harder although the C-P structure exists for those networks. On the other hand, 12 networks have cores that are much more influential than their periphery (i.e., $r_{c}/r_{p} >6$). 

%$(r_{c},\:\tilde{\rho}_{cc},\:\tilde{\rho}_{cp}) \approx (2.0904,\:10^{0.8},\:10^{0.4})$ is the marginal data for the relatively more significantly influential cores The data with $\tilde{\rho}_{cc} < 10^{0.8}$ or $\tilde{\rho}_{cc} < 10^{0.8}$ shows $r_{c}/r_{p} < 2$. and there are 44 networks satisfies the condition but their $r_{c}/r_{p}$ is less than or equal to $2$.

FIG.~\ref{fig:influence_pattern_real-world}. (b) demonstrates the increasing pattern of $r_{c}/r_{p}$ with $\tilde{\rho}_{cc}$ and $\tilde{\rho}_{cp}$. Although networks with $ 0 < r_{c}/r_{p} \leq 2$ mainly arise when $\tilde{\rho}_{cc} < 10^{1.5}$ and $\tilde{\rho}_{cp} < 10^{1.0}$, the increasing pattern is smooth rather than abrupt. The 12 networks with cores of most relative influence satisfy $\tilde{\rho}_{cc} > \tilde{\rho}_{cp}$. This observation shows that the relative influence of the core can be affected by the detailed features of C-P 
structures rather than the existence of the C-P structures themselves. 

Third, we investigated how the internal and external connectivity of the core are correlated to $r_{c}/r_{p}$. To do this, we demonstrate the patterns of $r_{c}/r_{p}$ with the internal$(\tilde{\rho}_{cc})$ and external$(\tilde{\rho}_{cp})$ connectivity. 

%we calculate Pearson's correlations for $(r_{c}/r_{p}, \: \tilde{\rho}_{cc})$ and $(r_{c}/r_{p}, \: \tilde{\rho}_{cp})$. 

FIG.~\ref{fig:influence_pattern_real-world}. (c) and (d) show how those two core's connectivities affect $r_{c}/r_{p}$. FIG.~\ref{fig:influence_pattern_real-world}. (c) shows $r_{c}/r_{p}$ is positively correlated to $\tilde{\rho}_{cc}$. The role of the internal connectivity increases the influence of core nodes before the spreading process ends by providing many routes inside the core. As a consequence, most of the core members can be affected by a single core node, and their connectivity to the periphery can increase the core node's influence further, making the core relatively more influential on average. 

FIG.~\ref{fig:influence_pattern_real-world}. (d) shows the relationship between $r_{c}/r_{p}$ and $\tilde{\rho}_{cp}$. The averaged $r_{c}/r_{p}$(the dashed line with triangle markers) shows that $r_{c}/r_{p}$ increases with $\tilde{\rho}_{cp}$ at the beginning, but it decreases where $\tilde{\rho}_{cp}$ is high. A low relative influence of the cores can be expected in an isolated core\cite{liu2015core} when $\tilde{\rho}_{cp}$ is closer to or less than $1(10^{0})$. However, our results show that while the relative influence of the core increases for moderate values of $\tilde{\rho}_{cp}$, this trend does not continue monotonically. Indeed, $r_c/r_p$ may even fall if $rho_cp$ gets too high.

However, our finding shows that a better external connectivity($\tilde{\rho}_{cp}$) does not always give a higher relative influence of the cores. High $\tilde{\rho}_{cp}$ can reduce $r_{c}/r_{p}$ if it is large. 

The Pearson's correlations of $(r_{c}/r_{p},\: \log_{10}(\tilde{\rho}_{cc}))$ and $(r_{c}/r_{p},\: \log_{10}(\tilde{\rho}_{cp}))$ show that the internal and external connectivity is significantly correlated to $r_{c}/r_{p}$ in a statistical sense with both $p$-values are less than $10^{-5}$. However, the correlation coefficient between $r_{c}/r_{p}$ and $\log_{10}(\tilde{\rho}_{cc})$ is about $1.35$ times larger than the coefficient between $r_{c}/r_{p}$ and $\log_{10}(\tilde{\rho}_{cp})$.

\subsection{Observation on simple 2-block C-P model }
\begin{figure}[ht]  
% \begin{minipage}{.5\linewidth}
  \centering
      \includegraphics[width=\linewidth]{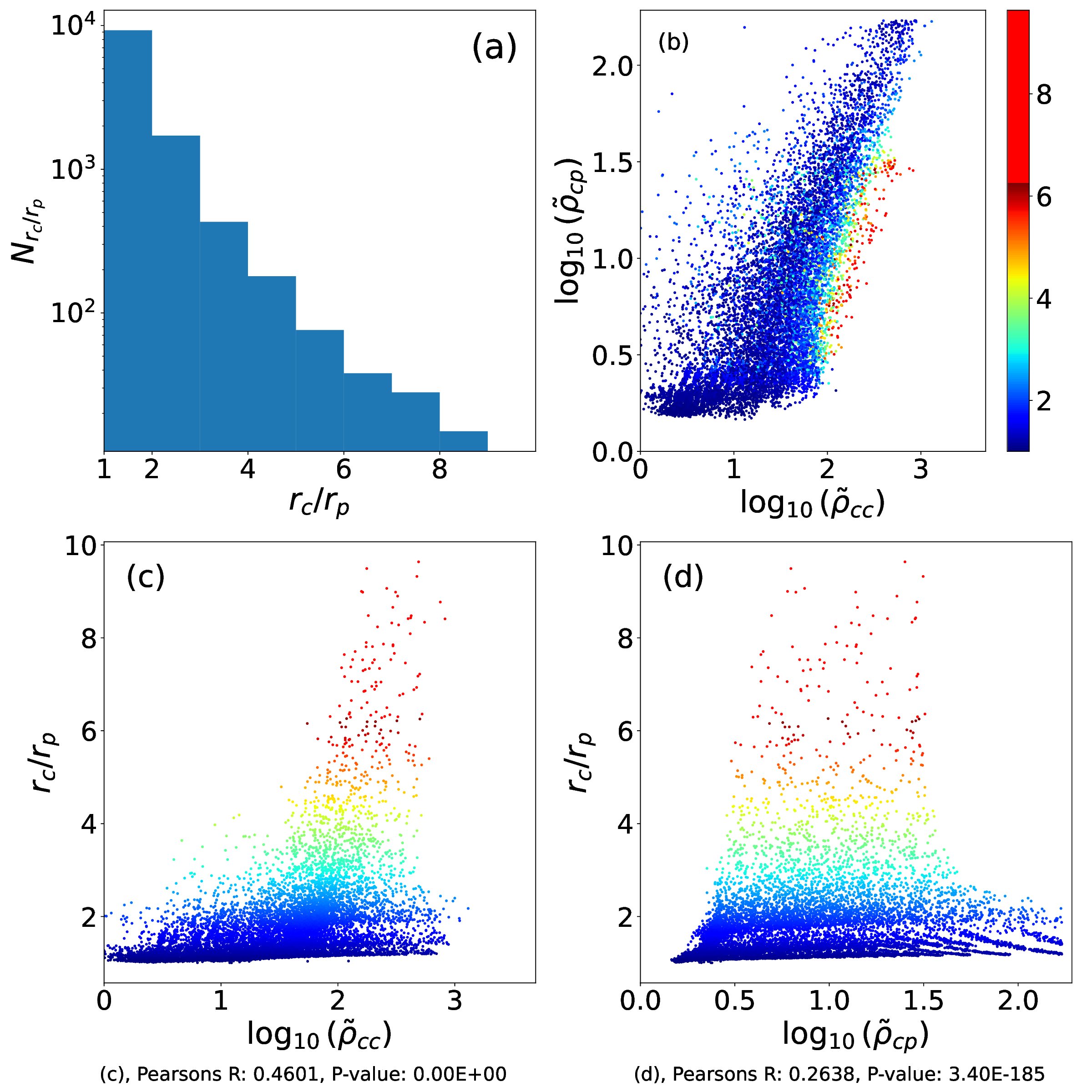}
  \caption{$r_{c}/r_{p}$ and its relationship to $(\tilde{\rho}_{cc}, \:\tilde{\rho}_{cp})$ in model-generated networks. The plot mirrors Fig.~\ref{fig:influence_pattern_real-world}. However, FIG.~\ref{fig:2-block-result}. (a) uses log base 10 on the y-axis due to the large scale. We include data where the most influential node can infect at least $10\%$ of the networks.} 
  \label{fig:2-block-result}
% \end{minipage} \hspace{5mm}
\end{figure}

% 앞에도 썼지만 모델을 보는 이유는 real-world network에는 C-P외에도 여러가지 구조적 특징이 있으니 C-P외에는 무작위적인 모델 네트워크를 고려해서 C-P 구조에 따라 core influence가 어떻게 달라지는지 좀 더 (제한적이지만) 명확히 보기 위해 모델을 고려했다고 써야하지 않을까요?

%The real-world data shows positive correlation between $\tilde{\rho}_{cc}$ and $\tilde{\rho}_{cp}$(FIG.~\ref{fig:corr_cla_intr}. (a)). The correlation can be a result of complex characteristics of real-world networks. Also, the correlation limits the range of $\tilde{\rho}_{cc}$ and $\tilde{\rho}_{cp}$.

%왼쪽 두 문단에서 모델의 결과가 real-net보다 더 넓은 rho_cc, rho_cp 영역에서 real-net에서 관측한 주요결과 2, 3을 뒷받침해준다는 핵심메시지가 잘 전달되고 있나요? 시간 부족으로 제가 자세히 쓰기가 힘드네요

We use model networks with a two-block structure to better understand the role of the internal and external connectivity of the cores. We generate $10$ model networks with fixed network size($N=4000$), core size($\alpha N = 0.1N$), and link densities$(\rho_{cc},\:\rho_{cp},\:\rho_{pp})$. We select the range of link densities that correspond to real-world data.

%FIG.~\ref{fig:2-block-result} shows broader range of $(\tilde{\rho}_{cc}, \:\tilde{\rho}_{cp})$ than the real-world case.

%in our selection

FIG.~\ref{fig:2-block-result}. (a) shows that the generated networks produce a range of $r_{c}/r_{p}$ from $1$ to $\sim 9$ despite the model's simplicity. FIG.~\ref{fig:2-block-result}. (b) shows that the model generates a broader range of $\tilde{\rho}_{cc}$ and $\tilde{\rho}_{cp}$ than the real-world networks. Since the model can generate networks with the broader magnitudes of the relative influence($r_{c}/r_{p}$) and the connectivity measures of the core$(\tilde{\rho}_{cc}, \: \tilde{\rho}_{cp})$, our data from the model can show the relationship between $r_{c}/r_{p}$ and $(\tilde{\rho}_{cc}, \: \tilde{\rho}_{cp})$ more clearly.
 
As observed in the real-world networks, FIG.~\ref{fig:2-block-result}. (b) shows that higher $r_{c}/r_{p}$ are associated with higher $(\tilde{\rho}_{cc} ,\:\tilde{\rho}_{cp})$. Furthermore, relatively influential cores of $r_{c}/r_{p} > 6$ can be found where the ratio  $\tilde{\rho}_{cc}/\tilde{\rho}_{cp} > 2 $, $\tilde{\rho}_{cc} > 10^{2}$, and $\tilde{\rho}_{cc} > 10^{0.5}$, which indicates the relatively more influential cores can be found in the ideal core with high enough internal and external connectivity of the core. Also, the smooth change of $r_{c}/r_{p}$ on $(\tilde{\rho}_{cc}, \: \tilde{\rho}_{cp})$ space supports the idea that the internal and external connectivity are correlated to the level of the relative influence of the core.

FIG.~\ref{fig:2-block-result}. (c) shows that $\tilde{\rho}_{cc}$ is positively correlated to $r_{c}/r_{p}$ as in the real-world results. Although $r_{c}/r_{p} < 2$ is more common with high $\tilde{\rho}_{cc}$, these correspond to networks with $\tilde{\rho}_{cc} \sim \tilde{\rho}_{cp}$.  

FIG.~\ref{fig:2-block-result}. (d) shows the optimal behaviour of $\tilde{\rho}_{cp}$ on $r_{c}/r_{p}$ more clearly than real-world networks. We infer that $r_{c}/r_{p}$ increases in line with the core when the internally well-connected core exists, but not when $\tilde{\rho}_{cp} \gtrsim 10^{1.8}$. The trend is much clearer than for real-world networks as we can isolate the effect of $\tilde{\rho}_{cp}$ (in real-world networks, both $\tilde{\rho}_{cc}$ and $\tilde{\rho}_{cp}$ can be large, masking the effect of the subdominant parameter). Indeed, it is very clear in this picture that very high values of $\tilde{\rho}_{cp}$ bring the relative influence of the core right down.

The Pearson's correlation coefficient for $\left( r_{c}/r_{p},\: \log_{10}(\tilde{\rho}_{cc}) \right)$ is $R \approx 0.4601$ with $p\text{-value} < 10^{-5}$, and the values for $\left(r_{c}/r_{p},\: \log_{10}(\tilde{\rho}_{cp}) \right)$ are $R \approx 0.2638$ with $p\text{-value} < 10^{-5}$. The higher correlation for $\tilde{\rho}_{cc}$ compared with $\tilde{\rho}_{cp}$ again follows the pattern seen in real-world data.

\section{\label{sec:Discussion I}Discussion}

This study looks to determine structural conditions for influential cores in networks in terms of the internal and external connectivity of the cores. Our findings show that those connectivities are distributed broadly in real-world networks, and they are positively correlated to the relative influence of the core. Influential cores are expected to be in ideal C-P structures, especially with relatively high internal and external connectivity of the core. Specifically the most influential cores($ r_c/r_p> 6$) are observed when cores are distinctly dense($ \tilde{\rho}_{cc}> 10^{1.0}$) and well-connected to the periphery($ \tilde{\rho}_{cp}> 10^{0.8}$) in real-world, and $ \tilde{\rho}_{cc}> 10^{2.0}$ and $ \tilde{\rho}_{cp}> 10^{0.5}$ in simple 2-block model networks, respectively. Furthermore, investigation of the Pearson's correlation shows that $\tilde{\rho}_{cc}$ is more highly correlated to $r_{c}/r_{p}$ than $\tilde{\rho}_{cp}$.

Generally, finding influential core nodes may require several conditions to be satisfied. The correlations we have found reflect the fact that higher internal connectivity should increase the relative influence of the core since the quantity is directly correlated to the number of routes for spreading influence inside the core\cite{kitsak2010identification}. More internal spreading means a higher influence on the core nodes since they tend to have reasonably enough external connections to the periphery. A lack of external connectivity can limit the influence of the core\cite{liu2015core}, so that is also a crucial factor for the core to be relatively more influential than the periphery. But there is a caveat to this statement. If the external connectivity of the core is too high, its relative influence can decrease. The underlying reason is that higher external connectivity allows more routes between the core and periphery. Since those extra routes increase the influence of the core and periphery simultaneously, the relative difference between the influence of the core and the periphery is reduced.

The Pearson's correlations between the relative influence of the core and each of the external and internal connectivity also support the idea that the internal connectivity is the most important factor in both the real-world and the generated networks. 

In short, the clearest sign of an influential core in a network is high internal connectivity and reasonably high external connectivity, and the internal connectivity is distinctly more pronounced than the external connectivity.  

%core => many route, periphery to core transition => speed gap, 본문에서 density 와 routes 를 연결했으므로 스피드 보자는 문제는 거낼 법도 하다.(데이터 확인해보자)

%In this work we have looked at the size of spread of an infection model to gauge influence. Further studies could look to see if speed is also an important factor, too.

%Further studies may investigate the influence of core in terms of infection speed since the influence of a node is not only about how far it can spread but also how quickly it can spread.

In this work, we have looked at the size of the spread of an infection model to gauge influence. Further studies could investigate the influence of core in terms of infection speed since the influence of a node is not only about how far it can spread but also how quickly it can spread. Previous studies have identified more complex C-P structures, such as layered or multiple cores\cite{zhang2015identification,kojaku2018core,gallagher2021clarified}. We avoid such complexity and capture just two structural features with the Lip method. Using more complicated features might help find influential nodes more accurately. For instance, in layered networks, we might think of the correlation between the core's influence and the internal connectivity of the layered feature. An extension of the suggested approach in this study may be beneficial for finding influential nodes, particularly with non-binary partitions such as the layered C-P. And there are questions to answer if there are multiple cores\cite{kojaku2017finding}. For instance, is the external connectivity between two cores more crucial than the external connectivity between the cores and periphery?

%[1]우리가 $\beta$ 에 대해 보여준게 별로 없기때문에 이 문단과 관련된 내용은 억측일 가능성이 높음. 빼는게 낫겠다. or... [2]그래도 이 연구에서 쓰인 $\beta$ 값 이외에 더 조사해보는게 낫다, 예컨데, structural features 에 따라 influence ratio 를 극대화 시킬 수 있는 $\beta$ 값이 존재할 수 있으며 이는 network structure optimisation 에 중요한 역할을 할 수도 있을 것이다. <= 이게 책임감 없는 이야기란거 

%We investigate various infection rates that fit a certain criterion, which requires the most influential node to infect $10\%$ of the entire network. This is a necessary condition to see how the location of the node is related to the influence in mesoscale structure. 

Our findings provide a structural condition for influence of cores in networks and a plausible explanation of why some cores are influential and others are not. We believe that our results are helpful in identifying influential node groups in networks by leveraging structural features that are simple to calculate.

\begin{acknowledgments}
% We wish to acknowledge the support of the author community in using
% REV\TeX{}, offering suggestions and encouragement, testing new versions,
% \dots.
Y.-H.E. acknowledge financial support by the National Research Foundation of Korea (NRF) grant funded by the Korea government (MSIT) (Grant No. 2020R1G1A1101950) and by the Basic Science Research Program through the National Research Foundation of Korea (NRF) funded by the Ministry of Education (Grant No. 2018R1A6A1A06024977).
\end{acknowledgments}

\appendix

\nocite{*}

\bibliography{apssamp.bib}% Produces the bibliography via BibTeX.

\begin{thebibliography}{46}%
\makeatletter
\providecommand \@ifxundefined [1]{%
 \@ifx{#1\undefined}
}%
\providecommand \@ifnum [1]{%
 \ifnum #1\expandafter \@firstoftwo
 \else \expandafter \@secondoftwo
 \fi
}%
\providecommand \@ifx [1]{%
 \ifx #1\expandafter \@firstoftwo
 \else \expandafter \@secondoftwo
 \fi
}%
\providecommand \natexlab [1]{#1}%
\providecommand \enquote  [1]{``#1''}%
\providecommand \bibnamefont  [1]{#1}%
\providecommand \bibfnamefont [1]{#1}%
\providecommand \citenamefont [1]{#1}%
\providecommand \href@noop [0]{\@secondoftwo}%
\providecommand \href [0]{\begingroup \@sanitize@url \@href}%
\providecommand \@href[1]{\@@startlink{#1}\@@href}%
\providecommand \@@href[1]{\endgroup#1\@@endlink}%
\providecommand \@sanitize@url [0]{\catcode `\\12\catcode `\$12\catcode `\&12\catcode `\#12\catcode `\^12\catcode `\_12\catcode `\%12\relax}%
\providecommand \@@startlink[1]{}%
\providecommand \@@endlink[0]{}%
\providecommand \url  [0]{\begingroup\@sanitize@url \@url }%
\providecommand \@url [1]{\endgroup\@href {#1}{\urlprefix }}%
\providecommand \urlprefix  [0]{URL }%
\providecommand \Eprint [0]{\href }%
\providecommand \doibase [0]{https://doi.org/}%
\providecommand \selectlanguage [0]{\@gobble}%
\providecommand \bibinfo  [0]{\@secondoftwo}%
\providecommand \bibfield  [0]{\@secondoftwo}%
\providecommand \translation [1]{[#1]}%
\providecommand \BibitemOpen [0]{}%
\providecommand \bibitemStop [0]{}%
\providecommand \bibitemNoStop [0]{.\EOS\space}%
\providecommand \EOS [0]{\spacefactor3000\relax}%
\providecommand \BibitemShut  [1]{\csname bibitem#1\endcsname}%
\let\auto@bib@innerbib\@empty
%</preamble>
\bibitem [{\citenamefont {Montes-Orozco}\ \emph {et~al.}(2022)\citenamefont {Montes-Orozco}, \citenamefont {Mora-Guti{\'e}rrez}, \citenamefont {de-los Cobos-Silva}, \citenamefont {Rinc{\'o}n-Garc{\'\i}a}, \citenamefont {Guti{\'e}rrez-Andrade}, \citenamefont {Lara-Vel{\'a}zquez} \emph {et~al.}}]{montes2022analysis}%
  \BibitemOpen
  \bibfield  {author} {\bibinfo {author} {\bibfnamefont {E.}~\bibnamefont {Montes-Orozco}}, \bibinfo {author} {\bibfnamefont {R.-A.}\ \bibnamefont {Mora-Guti{\'e}rrez}}, \bibinfo {author} {\bibfnamefont {S.-G.}\ \bibnamefont {de-los Cobos-Silva}}, \bibinfo {author} {\bibfnamefont {E.~A.}\ \bibnamefont {Rinc{\'o}n-Garc{\'\i}a}}, \bibinfo {author} {\bibfnamefont {M.~A.}\ \bibnamefont {Guti{\'e}rrez-Andrade}}, \bibinfo {author} {\bibfnamefont {P.}~\bibnamefont {Lara-Vel{\'a}zquez}}, \emph {et~al.},\ }\bibfield  {title} {\bibinfo {title} {Analysis and characterization of the spread of covid-19 in mexico through complex networks and optimization approaches},\ }\href@noop {} {\bibfield  {journal} {\bibinfo  {journal} {Complexity}\ }\textbf {\bibinfo {volume} {2022}} (\bibinfo {year} {2022})}\BibitemShut {NoStop}%
\bibitem [{\citenamefont {Meng}\ \emph {et~al.}(2022)\citenamefont {Meng}, \citenamefont {Han}, \citenamefont {Wu}, \citenamefont {Si},\ and\ \citenamefont {Cai}}]{meng2022analysis}%
  \BibitemOpen
  \bibfield  {author} {\bibinfo {author} {\bibfnamefont {X.}~\bibnamefont {Meng}}, \bibinfo {author} {\bibfnamefont {S.}~\bibnamefont {Han}}, \bibinfo {author} {\bibfnamefont {L.}~\bibnamefont {Wu}}, \bibinfo {author} {\bibfnamefont {S.}~\bibnamefont {Si}},\ and\ \bibinfo {author} {\bibfnamefont {Z.}~\bibnamefont {Cai}},\ }\bibfield  {title} {\bibinfo {title} {Analysis of epidemic vaccination strategies by node importance and evolutionary game on complex networks},\ }\href@noop {} {\bibfield  {journal} {\bibinfo  {journal} {Reliability Engineering \& System Safety}\ }\textbf {\bibinfo {volume} {219}},\ \bibinfo {pages} {108256} (\bibinfo {year} {2022})}\BibitemShut {NoStop}%
\bibitem [{\citenamefont {Zhang}\ \emph {et~al.}(2022)\citenamefont {Zhang}, \citenamefont {Li}, \citenamefont {Fan},\ and\ \citenamefont {Du}}]{zhang2022sei3r}%
  \BibitemOpen
  \bibfield  {author} {\bibinfo {author} {\bibfnamefont {Q.}~\bibnamefont {Zhang}}, \bibinfo {author} {\bibfnamefont {X.}~\bibnamefont {Li}}, \bibinfo {author} {\bibfnamefont {Y.}~\bibnamefont {Fan}},\ and\ \bibinfo {author} {\bibfnamefont {Y.}~\bibnamefont {Du}},\ }\bibfield  {title} {\bibinfo {title} {An sei3r information propagation control algorithm with structural hole and high influential infected nodes in social networks},\ }\href@noop {} {\bibfield  {journal} {\bibinfo  {journal} {Engineering Applications of Artificial Intelligence}\ }\textbf {\bibinfo {volume} {108}},\ \bibinfo {pages} {104573} (\bibinfo {year} {2022})}\BibitemShut {NoStop}%
\bibitem [{\citenamefont {Zareie}\ \emph {et~al.}(2019)\citenamefont {Zareie}, \citenamefont {Sheikhahmadi},\ and\ \citenamefont {Jalili}}]{zareie2019influential}%
  \BibitemOpen
  \bibfield  {author} {\bibinfo {author} {\bibfnamefont {A.}~\bibnamefont {Zareie}}, \bibinfo {author} {\bibfnamefont {A.}~\bibnamefont {Sheikhahmadi}},\ and\ \bibinfo {author} {\bibfnamefont {M.}~\bibnamefont {Jalili}},\ }\bibfield  {title} {\bibinfo {title} {Influential node ranking in social networks based on neighborhood diversity},\ }\href@noop {} {\bibfield  {journal} {\bibinfo  {journal} {Future Generation Computer Systems}\ }\textbf {\bibinfo {volume} {94}},\ \bibinfo {pages} {120} (\bibinfo {year} {2019})}\BibitemShut {NoStop}%
\bibitem [{\citenamefont {Ma}\ \emph {et~al.}(2016)\citenamefont {Ma}, \citenamefont {Ma}, \citenamefont {Zhang},\ and\ \citenamefont {Wang}}]{ma2016identifying}%
  \BibitemOpen
  \bibfield  {author} {\bibinfo {author} {\bibfnamefont {L.-l.}\ \bibnamefont {Ma}}, \bibinfo {author} {\bibfnamefont {C.}~\bibnamefont {Ma}}, \bibinfo {author} {\bibfnamefont {H.-F.}\ \bibnamefont {Zhang}},\ and\ \bibinfo {author} {\bibfnamefont {B.-H.}\ \bibnamefont {Wang}},\ }\bibfield  {title} {\bibinfo {title} {Identifying influential spreaders in complex networks based on gravity formula},\ }\href@noop {} {\bibfield  {journal} {\bibinfo  {journal} {Physica A: Statistical Mechanics and its Applications}\ }\textbf {\bibinfo {volume} {451}},\ \bibinfo {pages} {205} (\bibinfo {year} {2016})}\BibitemShut {NoStop}%
\bibitem [{\citenamefont {Liu}\ \emph {et~al.}(2016)\citenamefont {Liu}, \citenamefont {Tang}, \citenamefont {Zhou},\ and\ \citenamefont {Do}}]{liu2016identify}%
  \BibitemOpen
  \bibfield  {author} {\bibinfo {author} {\bibfnamefont {Y.}~\bibnamefont {Liu}}, \bibinfo {author} {\bibfnamefont {M.}~\bibnamefont {Tang}}, \bibinfo {author} {\bibfnamefont {T.}~\bibnamefont {Zhou}},\ and\ \bibinfo {author} {\bibfnamefont {Y.}~\bibnamefont {Do}},\ }\bibfield  {title} {\bibinfo {title} {Identify influential spreaders in complex networks, the role of neighborhood},\ }\href@noop {} {\bibfield  {journal} {\bibinfo  {journal} {Physica A: Statistical Mechanics and its Applications}\ }\textbf {\bibinfo {volume} {452}},\ \bibinfo {pages} {289} (\bibinfo {year} {2016})}\BibitemShut {NoStop}%
\bibitem [{\citenamefont {L{\"u}}\ \emph {et~al.}(2016{\natexlab{a}})\citenamefont {L{\"u}}, \citenamefont {Chen}, \citenamefont {Ren}, \citenamefont {Zhang}, \citenamefont {Zhang},\ and\ \citenamefont {Zhou}}]{lu2016vital}%
  \BibitemOpen
  \bibfield  {author} {\bibinfo {author} {\bibfnamefont {L.}~\bibnamefont {L{\"u}}}, \bibinfo {author} {\bibfnamefont {D.}~\bibnamefont {Chen}}, \bibinfo {author} {\bibfnamefont {X.-L.}\ \bibnamefont {Ren}}, \bibinfo {author} {\bibfnamefont {Q.-M.}\ \bibnamefont {Zhang}}, \bibinfo {author} {\bibfnamefont {Y.-C.}\ \bibnamefont {Zhang}},\ and\ \bibinfo {author} {\bibfnamefont {T.}~\bibnamefont {Zhou}},\ }\bibfield  {title} {\bibinfo {title} {Vital nodes identification in complex networks},\ }\href@noop {} {\bibfield  {journal} {\bibinfo  {journal} {Physics reports}\ }\textbf {\bibinfo {volume} {650}},\ \bibinfo {pages} {1} (\bibinfo {year} {2016}{\natexlab{a}})}\BibitemShut {NoStop}%
\bibitem [{\citenamefont {L{\"u}}\ \emph {et~al.}(2016{\natexlab{b}})\citenamefont {L{\"u}}, \citenamefont {Zhou}, \citenamefont {Zhang},\ and\ \citenamefont {Stanley}}]{lu2016h}%
  \BibitemOpen
  \bibfield  {author} {\bibinfo {author} {\bibfnamefont {L.}~\bibnamefont {L{\"u}}}, \bibinfo {author} {\bibfnamefont {T.}~\bibnamefont {Zhou}}, \bibinfo {author} {\bibfnamefont {Q.-M.}\ \bibnamefont {Zhang}},\ and\ \bibinfo {author} {\bibfnamefont {H.~E.}\ \bibnamefont {Stanley}},\ }\bibfield  {title} {\bibinfo {title} {The h-index of a network node and its relation to degree and coreness},\ }\href@noop {} {\bibfield  {journal} {\bibinfo  {journal} {Nature communications}\ }\textbf {\bibinfo {volume} {7}},\ \bibinfo {pages} {1} (\bibinfo {year} {2016}{\natexlab{b}})}\BibitemShut {NoStop}%
\bibitem [{\citenamefont {Hu}\ \emph {et~al.}(2013)\citenamefont {Hu}, \citenamefont {Gao}, \citenamefont {Ma}, \citenamefont {Yin}, \citenamefont {Zhang},\ and\ \citenamefont {Xing}}]{hu2013new}%
  \BibitemOpen
  \bibfield  {author} {\bibinfo {author} {\bibfnamefont {Q.}~\bibnamefont {Hu}}, \bibinfo {author} {\bibfnamefont {Y.}~\bibnamefont {Gao}}, \bibinfo {author} {\bibfnamefont {P.}~\bibnamefont {Ma}}, \bibinfo {author} {\bibfnamefont {Y.}~\bibnamefont {Yin}}, \bibinfo {author} {\bibfnamefont {Y.}~\bibnamefont {Zhang}},\ and\ \bibinfo {author} {\bibfnamefont {C.}~\bibnamefont {Xing}},\ }\bibfield  {title} {\bibinfo {title} {A new approach to identify influential spreaders in complex networks},\ }in\ \href@noop {} {\emph {\bibinfo {booktitle} {International Conference on Web-Age Information Management}}}\ (\bibinfo {organization} {Springer},\ \bibinfo {year} {2013})\ pp.\ \bibinfo {pages} {99--104}\BibitemShut {NoStop}%
\bibitem [{\citenamefont {Namtirtha}\ \emph {et~al.}(2021)\citenamefont {Namtirtha}, \citenamefont {Dutta}, \citenamefont {Dutta}, \citenamefont {Sundararajan},\ and\ \citenamefont {Simmhan}}]{namtirtha2021best}%
  \BibitemOpen
  \bibfield  {author} {\bibinfo {author} {\bibfnamefont {A.}~\bibnamefont {Namtirtha}}, \bibinfo {author} {\bibfnamefont {A.}~\bibnamefont {Dutta}}, \bibinfo {author} {\bibfnamefont {B.}~\bibnamefont {Dutta}}, \bibinfo {author} {\bibfnamefont {A.}~\bibnamefont {Sundararajan}},\ and\ \bibinfo {author} {\bibfnamefont {Y.}~\bibnamefont {Simmhan}},\ }\bibfield  {title} {\bibinfo {title} {Best influential spreaders identification using network global structural properties},\ }\href@noop {} {\bibfield  {journal} {\bibinfo  {journal} {Scientific reports}\ }\textbf {\bibinfo {volume} {11}},\ \bibinfo {pages} {2254} (\bibinfo {year} {2021})}\BibitemShut {NoStop}%
\bibitem [{\citenamefont {Borgatti}\ and\ \citenamefont {Everett}(2000)}]{borgatti2000models}%
  \BibitemOpen
  \bibfield  {author} {\bibinfo {author} {\bibfnamefont {S.~P.}\ \bibnamefont {Borgatti}}\ and\ \bibinfo {author} {\bibfnamefont {M.~G.}\ \bibnamefont {Everett}},\ }\bibfield  {title} {\bibinfo {title} {Models of core/periphery structures},\ }\href@noop {} {\bibfield  {journal} {\bibinfo  {journal} {Social networks}\ }\textbf {\bibinfo {volume} {21}},\ \bibinfo {pages} {375} (\bibinfo {year} {2000})}\BibitemShut {NoStop}%
\bibitem [{\citenamefont {Holme}(2005)}]{holme2005core}%
  \BibitemOpen
  \bibfield  {author} {\bibinfo {author} {\bibfnamefont {P.}~\bibnamefont {Holme}},\ }\bibfield  {title} {\bibinfo {title} {Core-periphery organization of complex networks},\ }\href@noop {} {\bibfield  {journal} {\bibinfo  {journal} {Physical Review E}\ }\textbf {\bibinfo {volume} {72}},\ \bibinfo {pages} {046111} (\bibinfo {year} {2005})}\BibitemShut {NoStop}%
\bibitem [{\citenamefont {Csermely}\ \emph {et~al.}(2013)\citenamefont {Csermely}, \citenamefont {London}, \citenamefont {Wu},\ and\ \citenamefont {Uzzi}}]{csermely2013structure}%
  \BibitemOpen
  \bibfield  {author} {\bibinfo {author} {\bibfnamefont {P.}~\bibnamefont {Csermely}}, \bibinfo {author} {\bibfnamefont {A.}~\bibnamefont {London}}, \bibinfo {author} {\bibfnamefont {L.-Y.}\ \bibnamefont {Wu}},\ and\ \bibinfo {author} {\bibfnamefont {B.}~\bibnamefont {Uzzi}},\ }\bibfield  {title} {\bibinfo {title} {Structure and dynamics of core/periphery networks},\ }\href@noop {} {\bibfield  {journal} {\bibinfo  {journal} {Journal of Complex Networks}\ }\textbf {\bibinfo {volume} {1}},\ \bibinfo {pages} {93} (\bibinfo {year} {2013})}\BibitemShut {NoStop}%
\bibitem [{\citenamefont {Kojaku}\ and\ \citenamefont {Masuda}(2018)}]{kojaku2018core}%
  \BibitemOpen
  \bibfield  {author} {\bibinfo {author} {\bibfnamefont {S.}~\bibnamefont {Kojaku}}\ and\ \bibinfo {author} {\bibfnamefont {N.}~\bibnamefont {Masuda}},\ }\bibfield  {title} {\bibinfo {title} {Core-periphery structure requires something else in the network},\ }\href@noop {} {\bibfield  {journal} {\bibinfo  {journal} {New Journal of Physics}\ }\textbf {\bibinfo {volume} {20}},\ \bibinfo {pages} {043012} (\bibinfo {year} {2018})}\BibitemShut {NoStop}%
\bibitem [{\citenamefont {Verma}\ \emph {et~al.}(2014)\citenamefont {Verma}, \citenamefont {Ara{\'u}jo},\ and\ \citenamefont {Herrmann}}]{verma2014revealing}%
  \BibitemOpen
  \bibfield  {author} {\bibinfo {author} {\bibfnamefont {T.}~\bibnamefont {Verma}}, \bibinfo {author} {\bibfnamefont {N.~A.}\ \bibnamefont {Ara{\'u}jo}},\ and\ \bibinfo {author} {\bibfnamefont {H.~J.}\ \bibnamefont {Herrmann}},\ }\bibfield  {title} {\bibinfo {title} {Revealing the structure of the world airline network},\ }\href@noop {} {\bibfield  {journal} {\bibinfo  {journal} {Scientific Reports}\ }\textbf {\bibinfo {volume} {4}},\ \bibinfo {pages} {1} (\bibinfo {year} {2014})}\BibitemShut {NoStop}%
\bibitem [{\citenamefont {Carmi}\ \emph {et~al.}(2007)\citenamefont {Carmi}, \citenamefont {Havlin}, \citenamefont {Kirkpatrick}, \citenamefont {Shavitt},\ and\ \citenamefont {Shir}}]{carmi2007model}%
  \BibitemOpen
  \bibfield  {author} {\bibinfo {author} {\bibfnamefont {S.}~\bibnamefont {Carmi}}, \bibinfo {author} {\bibfnamefont {S.}~\bibnamefont {Havlin}}, \bibinfo {author} {\bibfnamefont {S.}~\bibnamefont {Kirkpatrick}}, \bibinfo {author} {\bibfnamefont {Y.}~\bibnamefont {Shavitt}},\ and\ \bibinfo {author} {\bibfnamefont {E.}~\bibnamefont {Shir}},\ }\bibfield  {title} {\bibinfo {title} {A model of internet topology using k-shell decomposition},\ }\href@noop {} {\bibfield  {journal} {\bibinfo  {journal} {Proceedings of the National Academy of Sciences}\ }\textbf {\bibinfo {volume} {104}},\ \bibinfo {pages} {11150} (\bibinfo {year} {2007})}\BibitemShut {NoStop}%
\bibitem [{\citenamefont {Lee}\ \emph {et~al.}(2014)\citenamefont {Lee}, \citenamefont {Cucuringu},\ and\ \citenamefont {Porter}}]{lee2014density}%
  \BibitemOpen
  \bibfield  {author} {\bibinfo {author} {\bibfnamefont {S.~H.}\ \bibnamefont {Lee}}, \bibinfo {author} {\bibfnamefont {M.}~\bibnamefont {Cucuringu}},\ and\ \bibinfo {author} {\bibfnamefont {M.~A.}\ \bibnamefont {Porter}},\ }\bibfield  {title} {\bibinfo {title} {Density-based and transport-based core-periphery structures in networks},\ }\href@noop {} {\bibfield  {journal} {\bibinfo  {journal} {Physical Review E}\ }\textbf {\bibinfo {volume} {89}},\ \bibinfo {pages} {032810} (\bibinfo {year} {2014})}\BibitemShut {NoStop}%
\bibitem [{\citenamefont {Rombach}\ \emph {et~al.}(2014)\citenamefont {Rombach}, \citenamefont {Porter}, \citenamefont {Fowler},\ and\ \citenamefont {Mucha}}]{rombach2014core}%
  \BibitemOpen
  \bibfield  {author} {\bibinfo {author} {\bibfnamefont {M.~P.}\ \bibnamefont {Rombach}}, \bibinfo {author} {\bibfnamefont {M.~A.}\ \bibnamefont {Porter}}, \bibinfo {author} {\bibfnamefont {J.~H.}\ \bibnamefont {Fowler}},\ and\ \bibinfo {author} {\bibfnamefont {P.~J.}\ \bibnamefont {Mucha}},\ }\bibfield  {title} {\bibinfo {title} {Core-periphery structure in networks},\ }\href@noop {} {\bibfield  {journal} {\bibinfo  {journal} {SIAM Journal on Applied mathematics}\ }\textbf {\bibinfo {volume} {74}},\ \bibinfo {pages} {167} (\bibinfo {year} {2014})}\BibitemShut {NoStop}%
\bibitem [{\citenamefont {Kitsak}\ \emph {et~al.}(2010)\citenamefont {Kitsak}, \citenamefont {Gallos}, \citenamefont {Havlin}, \citenamefont {Liljeros}, \citenamefont {Muchnik}, \citenamefont {Stanley},\ and\ \citenamefont {Makse}}]{kitsak2010identification}%
  \BibitemOpen
  \bibfield  {author} {\bibinfo {author} {\bibfnamefont {M.}~\bibnamefont {Kitsak}}, \bibinfo {author} {\bibfnamefont {L.~K.}\ \bibnamefont {Gallos}}, \bibinfo {author} {\bibfnamefont {S.}~\bibnamefont {Havlin}}, \bibinfo {author} {\bibfnamefont {F.}~\bibnamefont {Liljeros}}, \bibinfo {author} {\bibfnamefont {L.}~\bibnamefont {Muchnik}}, \bibinfo {author} {\bibfnamefont {H.~E.}\ \bibnamefont {Stanley}},\ and\ \bibinfo {author} {\bibfnamefont {H.~A.}\ \bibnamefont {Makse}},\ }\bibfield  {title} {\bibinfo {title} {Identification of influential spreaders in complex networks},\ }\href@noop {} {\bibfield  {journal} {\bibinfo  {journal} {Nature Physics}\ }\textbf {\bibinfo {volume} {6}},\ \bibinfo {pages} {888} (\bibinfo {year} {2010})}\BibitemShut {NoStop}%
\bibitem [{\citenamefont {Bae}\ and\ \citenamefont {Kim}(2014)}]{bae2014identifying}%
  \BibitemOpen
  \bibfield  {author} {\bibinfo {author} {\bibfnamefont {J.}~\bibnamefont {Bae}}\ and\ \bibinfo {author} {\bibfnamefont {S.}~\bibnamefont {Kim}},\ }\bibfield  {title} {\bibinfo {title} {Identifying and ranking influential spreaders in complex networks by neighborhood coreness},\ }\href@noop {} {\bibfield  {journal} {\bibinfo  {journal} {Physica A: Statistical Mechanics and its Applications}\ }\textbf {\bibinfo {volume} {395}},\ \bibinfo {pages} {549} (\bibinfo {year} {2014})}\BibitemShut {NoStop}%
\bibitem [{\citenamefont {Liu}\ \emph {et~al.}(2015{\natexlab{a}})\citenamefont {Liu}, \citenamefont {Tang}, \citenamefont {Zhou},\ and\ \citenamefont {Do}}]{liu2015improving}%
  \BibitemOpen
  \bibfield  {author} {\bibinfo {author} {\bibfnamefont {Y.}~\bibnamefont {Liu}}, \bibinfo {author} {\bibfnamefont {M.}~\bibnamefont {Tang}}, \bibinfo {author} {\bibfnamefont {T.}~\bibnamefont {Zhou}},\ and\ \bibinfo {author} {\bibfnamefont {Y.}~\bibnamefont {Do}},\ }\bibfield  {title} {\bibinfo {title} {Improving the accuracy of the k-shell method by removing redundant links: From a perspective of spreading dynamics},\ }\href@noop {} {\bibfield  {journal} {\bibinfo  {journal} {Scientific reports}\ }\textbf {\bibinfo {volume} {5}},\ \bibinfo {pages} {1} (\bibinfo {year} {2015}{\natexlab{a}})}\BibitemShut {NoStop}%
\bibitem [{\citenamefont {Zhang}\ \emph {et~al.}(2016)\citenamefont {Zhang}, \citenamefont {Chen}, \citenamefont {Dong},\ and\ \citenamefont {Zhao}}]{zhang2016identifying}%
  \BibitemOpen
  \bibfield  {author} {\bibinfo {author} {\bibfnamefont {J.-X.}\ \bibnamefont {Zhang}}, \bibinfo {author} {\bibfnamefont {D.-B.}\ \bibnamefont {Chen}}, \bibinfo {author} {\bibfnamefont {Q.}~\bibnamefont {Dong}},\ and\ \bibinfo {author} {\bibfnamefont {Z.-D.}\ \bibnamefont {Zhao}},\ }\bibfield  {title} {\bibinfo {title} {Identifying a set of influential spreaders in complex networks},\ }\href@noop {} {\bibfield  {journal} {\bibinfo  {journal} {Scientific reports}\ }\textbf {\bibinfo {volume} {6}},\ \bibinfo {pages} {27823} (\bibinfo {year} {2016})}\BibitemShut {NoStop}%
\bibitem [{\citenamefont {Koujaku}\ \emph {et~al.}(2016)\citenamefont {Koujaku}, \citenamefont {Takigawa}, \citenamefont {Kudo},\ and\ \citenamefont {Imai}}]{koujaku2016dense}%
  \BibitemOpen
  \bibfield  {author} {\bibinfo {author} {\bibfnamefont {S.}~\bibnamefont {Koujaku}}, \bibinfo {author} {\bibfnamefont {I.}~\bibnamefont {Takigawa}}, \bibinfo {author} {\bibfnamefont {M.}~\bibnamefont {Kudo}},\ and\ \bibinfo {author} {\bibfnamefont {H.}~\bibnamefont {Imai}},\ }\bibfield  {title} {\bibinfo {title} {Dense core model for cohesive subgraph discovery},\ }\href@noop {} {\bibfield  {journal} {\bibinfo  {journal} {Social Networks}\ }\textbf {\bibinfo {volume} {44}},\ \bibinfo {pages} {143} (\bibinfo {year} {2016})}\BibitemShut {NoStop}%
\bibitem [{\citenamefont {Malliaros}\ \emph {et~al.}(2016)\citenamefont {Malliaros}, \citenamefont {Rossi},\ and\ \citenamefont {Vazirgiannis}}]{malliaros2016locating}%
  \BibitemOpen
  \bibfield  {author} {\bibinfo {author} {\bibfnamefont {F.~D.}\ \bibnamefont {Malliaros}}, \bibinfo {author} {\bibfnamefont {M.-E.~G.}\ \bibnamefont {Rossi}},\ and\ \bibinfo {author} {\bibfnamefont {M.}~\bibnamefont {Vazirgiannis}},\ }\bibfield  {title} {\bibinfo {title} {Locating influential nodes in complex networks},\ }\href@noop {} {\bibfield  {journal} {\bibinfo  {journal} {Scientific reports}\ }\textbf {\bibinfo {volume} {6}},\ \bibinfo {pages} {19307} (\bibinfo {year} {2016})}\BibitemShut {NoStop}%
\bibitem [{\citenamefont {Wang}\ \emph {et~al.}(2017)\citenamefont {Wang}, \citenamefont {Du}, \citenamefont {Fan},\ and\ \citenamefont {Xing}}]{wang2017ranking}%
  \BibitemOpen
  \bibfield  {author} {\bibinfo {author} {\bibfnamefont {Z.}~\bibnamefont {Wang}}, \bibinfo {author} {\bibfnamefont {C.}~\bibnamefont {Du}}, \bibinfo {author} {\bibfnamefont {J.}~\bibnamefont {Fan}},\ and\ \bibinfo {author} {\bibfnamefont {Y.}~\bibnamefont {Xing}},\ }\bibfield  {title} {\bibinfo {title} {Ranking influential nodes in social networks based on node position and neighborhood},\ }\href@noop {} {\bibfield  {journal} {\bibinfo  {journal} {Neurocomputing}\ }\textbf {\bibinfo {volume} {260}},\ \bibinfo {pages} {466} (\bibinfo {year} {2017})}\BibitemShut {NoStop}%
\bibitem [{\citenamefont {Liu}\ \emph {et~al.}(2015{\natexlab{b}})\citenamefont {Liu}, \citenamefont {Tang}, \citenamefont {Zhou},\ and\ \citenamefont {Do}}]{liu2015core}%
  \BibitemOpen
  \bibfield  {author} {\bibinfo {author} {\bibfnamefont {Y.}~\bibnamefont {Liu}}, \bibinfo {author} {\bibfnamefont {M.}~\bibnamefont {Tang}}, \bibinfo {author} {\bibfnamefont {T.}~\bibnamefont {Zhou}},\ and\ \bibinfo {author} {\bibfnamefont {Y.}~\bibnamefont {Do}},\ }\bibfield  {title} {\bibinfo {title} {Core-like groups result in invalidation of identifying super-spreader by k-shell decomposition},\ }\href@noop {} {\bibfield  {journal} {\bibinfo  {journal} {Scientific Reports}\ }\textbf {\bibinfo {volume} {5}},\ \bibinfo {pages} {1} (\bibinfo {year} {2015}{\natexlab{b}})}\BibitemShut {NoStop}%
\bibitem [{\citenamefont {Zhang}\ \emph {et~al.}(2015)\citenamefont {Zhang}, \citenamefont {Martin},\ and\ \citenamefont {Newman}}]{zhang2015identification}%
  \BibitemOpen
  \bibfield  {author} {\bibinfo {author} {\bibfnamefont {X.}~\bibnamefont {Zhang}}, \bibinfo {author} {\bibfnamefont {T.}~\bibnamefont {Martin}},\ and\ \bibinfo {author} {\bibfnamefont {M.~E.}\ \bibnamefont {Newman}},\ }\bibfield  {title} {\bibinfo {title} {Identification of core-periphery structure in networks},\ }\href@noop {} {\bibfield  {journal} {\bibinfo  {journal} {Physical Review E}\ }\textbf {\bibinfo {volume} {91}},\ \bibinfo {pages} {032803} (\bibinfo {year} {2015})}\BibitemShut {NoStop}%
\bibitem [{\citenamefont {Gallagher}\ \emph {et~al.}(2021)\citenamefont {Gallagher}, \citenamefont {Young},\ and\ \citenamefont {Welles}}]{gallagher2021clarified}%
  \BibitemOpen
  \bibfield  {author} {\bibinfo {author} {\bibfnamefont {R.~J.}\ \bibnamefont {Gallagher}}, \bibinfo {author} {\bibfnamefont {J.-G.}\ \bibnamefont {Young}},\ and\ \bibinfo {author} {\bibfnamefont {B.~F.}\ \bibnamefont {Welles}},\ }\bibfield  {title} {\bibinfo {title} {A clarified typology of core-periphery structure in networks},\ }\href@noop {} {\bibfield  {journal} {\bibinfo  {journal} {Science advances}\ }\textbf {\bibinfo {volume} {7}},\ \bibinfo {pages} {eabc9800} (\bibinfo {year} {2021})}\BibitemShut {NoStop}%
\bibitem [{\citenamefont {Lip}(2011)}]{lip2011fast}%
  \BibitemOpen
  \bibfield  {author} {\bibinfo {author} {\bibfnamefont {S.~Z.}\ \bibnamefont {Lip}},\ }\bibfield  {title} {\bibinfo {title} {A fast algorithm for the discrete core/periphery bipartitioning problem},\ }\href@noop {} {\bibfield  {journal} {\bibinfo  {journal} {arXiv preprint arXiv:1102.5511}\ } (\bibinfo {year} {2011})}\BibitemShut {NoStop}%
\bibitem [{\citenamefont {Bargigli}\ \emph {et~al.}(2015)\citenamefont {Bargigli}, \citenamefont {Di~Iasio}, \citenamefont {Infante}, \citenamefont {Lillo},\ and\ \citenamefont {Pierobon}}]{bargigli2015multiplex}%
  \BibitemOpen
  \bibfield  {author} {\bibinfo {author} {\bibfnamefont {L.}~\bibnamefont {Bargigli}}, \bibinfo {author} {\bibfnamefont {G.}~\bibnamefont {Di~Iasio}}, \bibinfo {author} {\bibfnamefont {L.}~\bibnamefont {Infante}}, \bibinfo {author} {\bibfnamefont {F.}~\bibnamefont {Lillo}},\ and\ \bibinfo {author} {\bibfnamefont {F.}~\bibnamefont {Pierobon}},\ }\bibfield  {title} {\bibinfo {title} {The multiplex structure of interbank networks},\ }\href@noop {} {\bibfield  {journal} {\bibinfo  {journal} {Quantitative Finance}\ }\textbf {\bibinfo {volume} {15}},\ \bibinfo {pages} {673} (\bibinfo {year} {2015})}\BibitemShut {NoStop}%
\bibitem [{\citenamefont {Lux}(2015)}]{lux2015emergence}%
  \BibitemOpen
  \bibfield  {author} {\bibinfo {author} {\bibfnamefont {T.}~\bibnamefont {Lux}},\ }\bibfield  {title} {\bibinfo {title} {Emergence of a core-periphery structure in a simple dynamic model of the interbank market},\ }\href@noop {} {\bibfield  {journal} {\bibinfo  {journal} {Journal of Economic Dynamics and Control}\ }\textbf {\bibinfo {volume} {52}},\ \bibinfo {pages} {A11} (\bibinfo {year} {2015})}\BibitemShut {NoStop}%
\bibitem [{\citenamefont {Barucca}\ \emph {et~al.}(2016)\citenamefont {Barucca}, \citenamefont {Tantari},\ and\ \citenamefont {Lillo}}]{barucca2016centrality}%
  \BibitemOpen
  \bibfield  {author} {\bibinfo {author} {\bibfnamefont {P.}~\bibnamefont {Barucca}}, \bibinfo {author} {\bibfnamefont {D.}~\bibnamefont {Tantari}},\ and\ \bibinfo {author} {\bibfnamefont {F.}~\bibnamefont {Lillo}},\ }\bibfield  {title} {\bibinfo {title} {Centrality metrics and localization in core-periphery networks},\ }\href@noop {} {\bibfield  {journal} {\bibinfo  {journal} {Journal of Statistical Mechanics: Theory and Experiment}\ }\textbf {\bibinfo {volume} {2016}},\ \bibinfo {pages} {023401} (\bibinfo {year} {2016})}\BibitemShut {NoStop}%
\bibitem [{\citenamefont {Karlov{\v{c}}ec}\ \emph {et~al.}(2016)\citenamefont {Karlov{\v{c}}ec}, \citenamefont {Lu{\v{z}}ar},\ and\ \citenamefont {Mladeni{\'c}}}]{karlovvcec2016core}%
  \BibitemOpen
  \bibfield  {author} {\bibinfo {author} {\bibfnamefont {M.}~\bibnamefont {Karlov{\v{c}}ec}}, \bibinfo {author} {\bibfnamefont {B.}~\bibnamefont {Lu{\v{z}}ar}},\ and\ \bibinfo {author} {\bibfnamefont {D.}~\bibnamefont {Mladeni{\'c}}},\ }\bibfield  {title} {\bibinfo {title} {Core-periphery dynamics in collaboration networks: the case study of slovenia},\ }\href@noop {} {\bibfield  {journal} {\bibinfo  {journal} {Scientometrics}\ }\textbf {\bibinfo {volume} {109}},\ \bibinfo {pages} {1561} (\bibinfo {year} {2016})}\BibitemShut {NoStop}%
\bibitem [{\citenamefont {Alvarez-Hamelin}\ \emph {et~al.}(2005)\citenamefont {Alvarez-Hamelin}, \citenamefont {Dall'Asta}, \citenamefont {Barrat},\ and\ \citenamefont {Vespignani}}]{alvarez2005k}%
  \BibitemOpen
  \bibfield  {author} {\bibinfo {author} {\bibfnamefont {J.~I.}\ \bibnamefont {Alvarez-Hamelin}}, \bibinfo {author} {\bibfnamefont {L.}~\bibnamefont {Dall'Asta}}, \bibinfo {author} {\bibfnamefont {A.}~\bibnamefont {Barrat}},\ and\ \bibinfo {author} {\bibfnamefont {A.}~\bibnamefont {Vespignani}},\ }\bibfield  {title} {\bibinfo {title} {k-core decomposition: A tool for the visualization of large scale networks},\ }\href@noop {} {\bibfield  {journal} {\bibinfo  {journal} {arXiv preprint cs/0504107}\ } (\bibinfo {year} {2005})}\BibitemShut {NoStop}%
\bibitem [{\citenamefont {Ure{\~n}a-Carri{\'o}n}\ \emph {et~al.}(2023)\citenamefont {Ure{\~n}a-Carri{\'o}n}, \citenamefont {Karimi}, \citenamefont {{\'I}{\~n}iguez},\ and\ \citenamefont {Kivel{\"a}}}]{urena2023assortative}%
  \BibitemOpen
  \bibfield  {author} {\bibinfo {author} {\bibfnamefont {J.}~\bibnamefont {Ure{\~n}a-Carri{\'o}n}}, \bibinfo {author} {\bibfnamefont {F.}~\bibnamefont {Karimi}}, \bibinfo {author} {\bibfnamefont {G.}~\bibnamefont {{\'I}{\~n}iguez}},\ and\ \bibinfo {author} {\bibfnamefont {M.}~\bibnamefont {Kivel{\"a}}},\ }\bibfield  {title} {\bibinfo {title} {Assortative and preferential attachment lead to core-periphery networks},\ }\href@noop {} {\bibfield  {journal} {\bibinfo  {journal} {Physical Review Research}\ }\textbf {\bibinfo {volume} {5}},\ \bibinfo {pages} {043287} (\bibinfo {year} {2023})}\BibitemShut {NoStop}%
\bibitem [{\citenamefont {Pastor-Satorras}\ \emph {et~al.}(2015)\citenamefont {Pastor-Satorras}, \citenamefont {Castellano}, \citenamefont {Van~Mieghem},\ and\ \citenamefont {Vespignani}}]{pastor2015epidemic}%
  \BibitemOpen
  \bibfield  {author} {\bibinfo {author} {\bibfnamefont {R.}~\bibnamefont {Pastor-Satorras}}, \bibinfo {author} {\bibfnamefont {C.}~\bibnamefont {Castellano}}, \bibinfo {author} {\bibfnamefont {P.}~\bibnamefont {Van~Mieghem}},\ and\ \bibinfo {author} {\bibfnamefont {A.}~\bibnamefont {Vespignani}},\ }\bibfield  {title} {\bibinfo {title} {Epidemic processes in complex networks},\ }\href@noop {} {\bibfield  {journal} {\bibinfo  {journal} {Reviews of modern physics}\ }\textbf {\bibinfo {volume} {87}},\ \bibinfo {pages} {925} (\bibinfo {year} {2015})}\BibitemShut {NoStop}%
\bibitem [{\citenamefont {Min}(2018)}]{min2018identifying}%
  \BibitemOpen
  \bibfield  {author} {\bibinfo {author} {\bibfnamefont {B.}~\bibnamefont {Min}},\ }\bibfield  {title} {\bibinfo {title} {Identifying an influential spreader from a single seed in complex networks via a message-passing approach},\ }\href@noop {} {\bibfield  {journal} {\bibinfo  {journal} {The European Physical Journal B}\ }\textbf {\bibinfo {volume} {91}},\ \bibinfo {pages} {1} (\bibinfo {year} {2018})}\BibitemShut {NoStop}%
\bibitem [{\citenamefont {Erdos}\ \emph {et~al.}(1960)\citenamefont {Erdos}, \citenamefont {R{\'e}nyi} \emph {et~al.}}]{erdos1960evolution}%
  \BibitemOpen
  \bibfield  {author} {\bibinfo {author} {\bibfnamefont {P.}~\bibnamefont {Erdos}}, \bibinfo {author} {\bibfnamefont {A.}~\bibnamefont {R{\'e}nyi}}, \emph {et~al.},\ }\bibfield  {title} {\bibinfo {title} {On the evolution of random graphs},\ }\href@noop {} {\bibfield  {journal} {\bibinfo  {journal} {Publ. math. inst. hung. acad. sci}\ }\textbf {\bibinfo {volume} {5}},\ \bibinfo {pages} {17} (\bibinfo {year} {1960})}\BibitemShut {NoStop}%
\bibitem [{\citenamefont {Peixoto}(2020)}]{Netzsch}%
  \BibitemOpen
  \bibfield  {author} {\bibinfo {author} {\bibfnamefont {T.~P.}\ \bibnamefont {Peixoto}},\ }\bibfield  {title} {\bibinfo {title} {The netzschleuder network catalogue and repository}\ }(\bibinfo {year} {2020})\BibitemShut {NoStop}%
\bibitem [{\citenamefont {Kunegis}(2013)}]{konect}%
  \BibitemOpen
  \bibfield  {author} {\bibinfo {author} {\bibfnamefont {J.}~\bibnamefont {Kunegis}},\ }\bibfield  {title} {\bibinfo {title} {{KONECT} -- {The} {Koblenz} {Network} {Collection}},\ }in\ \href {http://dl.acm.org/citation.cfm?id=2488173} {\emph {\bibinfo {booktitle} {Proc. Int. Conf. on World Wide Web Companion}}}\ (\bibinfo {year} {2013})\ pp.\ \bibinfo {pages} {1343--1350}\BibitemShut {NoStop}%
\bibitem [{\citenamefont {Kojaku}\ and\ \citenamefont {Masuda}(2017)}]{kojaku2017finding}%
  \BibitemOpen
  \bibfield  {author} {\bibinfo {author} {\bibfnamefont {S.}~\bibnamefont {Kojaku}}\ and\ \bibinfo {author} {\bibfnamefont {N.}~\bibnamefont {Masuda}},\ }\bibfield  {title} {\bibinfo {title} {Finding multiple core-periphery pairs in networks},\ }\href@noop {} {\bibfield  {journal} {\bibinfo  {journal} {Physical Review E}\ }\textbf {\bibinfo {volume} {96}},\ \bibinfo {pages} {052313} (\bibinfo {year} {2017})}\BibitemShut {NoStop}%
\bibitem [{\citenamefont {Kempe}\ \emph {et~al.}(2003)\citenamefont {Kempe}, \citenamefont {Kleinberg},\ and\ \citenamefont {Tardos}}]{kempe2003maximizing}%
  \BibitemOpen
  \bibfield  {author} {\bibinfo {author} {\bibfnamefont {D.}~\bibnamefont {Kempe}}, \bibinfo {author} {\bibfnamefont {J.}~\bibnamefont {Kleinberg}},\ and\ \bibinfo {author} {\bibfnamefont {{\'E}.}~\bibnamefont {Tardos}},\ }\bibfield  {title} {\bibinfo {title} {Maximizing the spread of influence through a social network},\ }in\ \href@noop {} {\emph {\bibinfo {booktitle} {Proceedings of the ninth ACM SIGKDD international conference on Knowledge discovery and data mining}}}\ (\bibinfo {year} {2003})\ pp.\ \bibinfo {pages} {137--146}\BibitemShut {NoStop}%
\bibitem [{\citenamefont {Wei}\ \emph {et~al.}(2022)\citenamefont {Wei}, \citenamefont {Zhao}, \citenamefont {Liu},\ and\ \citenamefont {Wang}}]{wei2022identifying}%
  \BibitemOpen
  \bibfield  {author} {\bibinfo {author} {\bibfnamefont {X.}~\bibnamefont {Wei}}, \bibinfo {author} {\bibfnamefont {J.}~\bibnamefont {Zhao}}, \bibinfo {author} {\bibfnamefont {S.}~\bibnamefont {Liu}},\ and\ \bibinfo {author} {\bibfnamefont {Y.}~\bibnamefont {Wang}},\ }\bibfield  {title} {\bibinfo {title} {Identifying influential spreaders in complex networks for disease spread and control},\ }\href@noop {} {\bibfield  {journal} {\bibinfo  {journal} {Scientific reports}\ }\textbf {\bibinfo {volume} {12}},\ \bibinfo {pages} {5550} (\bibinfo {year} {2022})}\BibitemShut {NoStop}%
\bibitem [{\citenamefont {Goyal}\ \emph {et~al.}(2013)\citenamefont {Goyal}, \citenamefont {Bonchi}, \citenamefont {Lakshmanan},\ and\ \citenamefont {Venkatasubramanian}}]{goyal2013minimizing}%
  \BibitemOpen
  \bibfield  {author} {\bibinfo {author} {\bibfnamefont {A.}~\bibnamefont {Goyal}}, \bibinfo {author} {\bibfnamefont {F.}~\bibnamefont {Bonchi}}, \bibinfo {author} {\bibfnamefont {L.~V.}\ \bibnamefont {Lakshmanan}},\ and\ \bibinfo {author} {\bibfnamefont {S.}~\bibnamefont {Venkatasubramanian}},\ }\bibfield  {title} {\bibinfo {title} {On minimizing budget and time in influence propagation over social networks},\ }\href@noop {} {\bibfield  {journal} {\bibinfo  {journal} {Social network analysis and mining}\ }\textbf {\bibinfo {volume} {3}},\ \bibinfo {pages} {179} (\bibinfo {year} {2013})}\BibitemShut {NoStop}%
\bibitem [{\citenamefont {Shen}\ \emph {et~al.}(2021)\citenamefont {Shen}, \citenamefont {Aliko}, \citenamefont {Han}, \citenamefont {Skipper},\ and\ \citenamefont {Peng}}]{shen2021finding}%
  \BibitemOpen
  \bibfield  {author} {\bibinfo {author} {\bibfnamefont {X.}~\bibnamefont {Shen}}, \bibinfo {author} {\bibfnamefont {S.}~\bibnamefont {Aliko}}, \bibinfo {author} {\bibfnamefont {Y.}~\bibnamefont {Han}}, \bibinfo {author} {\bibfnamefont {J.~I.}\ \bibnamefont {Skipper}},\ and\ \bibinfo {author} {\bibfnamefont {C.}~\bibnamefont {Peng}},\ }\bibfield  {title} {\bibinfo {title} {Finding core-periphery structures with node influences},\ }\href@noop {} {\bibfield  {journal} {\bibinfo  {journal} {IEEE Transactions on Network Science and Engineering}\ }\textbf {\bibinfo {volume} {9}},\ \bibinfo {pages} {875} (\bibinfo {year} {2021})}\BibitemShut {NoStop}%
\bibitem [{\citenamefont {Barab{\'a}si}\ \emph {et~al.}(2016)\citenamefont {Barab{\'a}si} \emph {et~al.}}]{barabasi2016network}%
  \BibitemOpen
  \bibfield  {author} {\bibinfo {author} {\bibfnamefont {A.-L.}\ \bibnamefont {Barab{\'a}si}} \emph {et~al.},\ }\href@noop {} {\emph {\bibinfo {title} {Network Science}}}\ (\bibinfo  {publisher} {Cambridge University Press},\ \bibinfo {year} {2016})\BibitemShut {NoStop}%
\end{thebibliography}%

\end{document}